%% file: main.tex
\documentclass{IEEEoj}
\usepackage{cite}
\usepackage{amsmath,amssymb,amsfonts}
\usepackage{algorithmic}
\usepackage{graphicx,color}
\usepackage{textcomp}
\def\BibTeX{{\rm B\kern-.05em{\sc i\kern-.025em b}\kern-.08em
    T\kern-.1667em\lower.7ex\hbox{E}\kern-.125emX}}
\AtBeginDocument{\definecolor{ojcolor}{cmyk}{0.93,0.59,0.15,0.02}}
\usepackage{amsmath,graphicx}
\usepackage{paralist}
\usepackage{multirow}
\usepackage{flushend}
\DeclareMathOperator*{\argmax}{arg\,max}
\usepackage[norelsize, linesnumbered, ruled, lined, boxed, commentsnumbered]{algorithm2e}
\usepackage{booktabs}
\usepackage{graphicx}
\usepackage{hyperref}

\begin{document}
\receiveddate{XX Month, XXXX}
\reviseddate{XX Month, XXXX}
\accepteddate{XX Month, XXXX}
\publisheddate{XX Month, XXXX}
\currentdate{XX Month, XXXX}

\title{Audio-Visual Activity Guided Cross-Modal Identity Association for Active Speaker Detection}

\author{RAHUL SHARMA, SHRIKANTH NARAYANAN, FELLOW IEEE}
\affil{University of Southern California, Los Angeles, CA 90007 USA}
\corresp{CORRESPONDING AUTHOR: Rahul Sharma (e-mail: rahul.sharma@usc.edu)}

\begin{abstract}
Active speaker detection in videos addresses associating a source face, visible in the video frames, with the underlying speech in the audio modality. The two primary sources of information to derive such a speech-face relationship are i) visual activity and its interaction with the speech signal and ii) co-occurrences of speakers' identities across modalities in the form of face and speech. The two approaches have their limitations: the audio-visual activity models get confused with other frequently occurring vocal activities, such as laughing and chewing, while the speakers' identity-based methods are limited to videos having enough disambiguating information to establish a speech-face association. Since the two approaches are independent, we investigate their complementary nature in this work. We propose a novel unsupervised framework to guide the speakers' cross-modal identity association with the audio-visual activity for active speaker detection.  Through experiments on entertainment media videos from two benchmark datasets––the AVA active speaker (movies) and Visual Person Clustering Dataset (TV shows)––we show that a simple late fusion of the two approaches enhances the active speaker detection performance. Implementation will be available: \href{https://github.com/usc-sail/SCMIA-unsupervised-ASD.git}{https://github.com/usc-sail/SCMIA-unsupervised-ASD.git}
\end{abstract}

\begin{IEEEkeywords}
Active speaker detection, face recognition, speaker recognition, character identity, comprehensive ASD, cross-modal
\end{IEEEkeywords}


\maketitle

\input{intro}

\input{methods}

\input{expt}

\section{CONCLUSION}
This work showed that the errors of the two independent systems for ASD, using: i) audio-visual activity and ii) co-occurrence of speakers' identity, have an exclusive component, validating the need to develop methods collaborating the two approaches. We proposed a generalized framework that can incorporate external information from any source while studying the co-occurrences of speakers' identities across modalities to establish speech-face associations. We have shown that its integration with fully supervised (TalkNet) and self-supervised (Syncnet) av-activity systems improve performance. As the framework is unsupervised, its integration with self-supervised av-activity eliminates the need for ASD labels. One immediate future direction is to enable speech-body association, which can help further enhance ASD, particularly helping cases where faces are in extreme poses or not visible. 

\newpage
\bibliographystyle{IEEEtran}
\bibliography{references}
\end{document}

%% file: intro.tex
\section{INTRODUCTION}
\IEEEPARstart{A}{ctive} speaker detection (ASD) aims at finding a source face corresponding to the speech activity present in the audio modality. It targets selecting a face from the set of possible candidate faces appearing in the video frames, such that the selected face and the foreground speech belong to the same person. Active speaker detection plays a fundamental building block in our target domain of Computational media intelligence~\cite{somandepalli2021computational}, which includes the study of character portrayals in media and their impact on society. The tools include speaker/character diarization~\cite{sharma2022using, intel_diariz}, audio-visual speech recognition~\cite{asr_google},  background character detection~\cite{sharma2022audio}, character role detection~\cite{deepstar}, speaker behavior understanding~\cite{sharma_publicspeaker}, etc.~. In this work, we address ASD in the context of entertainment media videos. These videos pose additional challenges to ASD due to the presence of an unknown and varying number of speakers and their highly dynamic interactions leading to faces in extreme poses. Speech present with noise and music further adds to the challenges. 

Focused on finding an association between speech and faces, ASD is inherently a multimodal task, requiring integration of the audio and visual information. The majority of previous research is situated around modeling the activity in the visual frames, in the form of visual actions, and activity in the audio modality, in the form of speech, and studying their interactions to establish speech-face associations~\cite{owens2018audio,bendris2010lip, arandjelovic2018objects, chung2016out, chung2019naver}. Earlier approaches have modeled the visual activity in the explicitly extracted mouth regions of faces and analyzed their correlations with the underlying speech waveforms to detect the active speaker faces~\cite{bendris2010lip}. More recently, researchers have proposed audio-visual synchronization as a proxy task to jointly model audio-visual activity in a self-supervised manner~\cite{chung2016out}. They showed its applicability to sound-source localization~\cite{owens2018audio}, which directly extends to ASD, as active speakers are essentially the sound sources for the underlying speech sound event. This led to various self-supervised methods, introducing several proxy tasks in multimodal~\cite{owens2018audio, arandjelovic2018objects, zhao2018sound, afouras2020self} and cross-modal~\cite{sharma2020cross, chakravarty2015s, chakravarty2016cross, sharma2019icip} scenarios targeted to active speaker detection and, sound source localization in general. 

Recently, several large-scale datasets consisting of face box-wise active speaker annotations for videos from movies~\cite{roth2020ava, kim2021look} and TV shows~\cite{brown2021face} have emerged. These datasets have led to various fully-supervised methods, where researchers model the audio-visual activity using sophisticated neural networks consisting of 3D CNNs and RNNs~\cite{chung2019naver, huang2020improved}. With further advancements in deep neural networks, researchers have modeled the audio-visual activity's short-term and long-term temporal dynamics using graph-convolutional neural networks and self-attention neural networks~\cite{alcazar2020active, leon2021maas, tao2021someone}. Researchers have also proposed to utilize context in the form of visual activity of the co-occurring speakers to enhance the performance of ASD further~\cite{min2022learning}. 

The recent advancements in face recognition~\cite{senet, guo2016ms} and speaker recognition (using speech in the audio modality) systems~\cite{indefense, voxceleb2}     have enabled the use of the speaker's identity present in the speech and the speaker's face to establish speech-face associations~\cite{hoover2018using, sharma2022unsupervised}. The idea of analyzing the co-occurrences of speakers' identities in the audio and the visual modalities to establish speech-face associations is not as well explored as modeling the visual activity for active speaker detection. Hoover et al.~\cite{hoover2018using} proposed to cluster the speech and faces independently and then used the temporal co-occurrence of the obtained clusters to establish speech-face correspondences. In one of our previous works~\cite{sharma2022unsupervised}, we constructed the temporal identity structure by observing the speech segments through the video. We proposed to select active speaker faces for each speech segment from the set of faces concurrently appearing in the frames, such that the chosen active speaker faces display similar identity structure as those obtained observing the speech segments.

The approaches to ASD mentioned above can be categorized based on the source of information into: i) audio-visual activity: the ASD relevant information is derived from modeling the audio and visual activities and their interactions. ii) speakers' identity co-occurrence: the speech-face correspondences are solved by studying the cross-modal co-occurrences of the speakers' identity (speech and faces). The audio-visual activity-based methods model the visual actions relevant to the speaking activity. In daily life scenarios, which are often present in entertainment media videos, these methods get confused with other actions similar to the speaking activity, such as eating, grinning, etc. They often miss the cases when the speaker's face appears in extreme poses and the lip region is not visible. On the other hand, the speakers' identity-based methods are independent of the observed visual actions while limited to videos having enough information to disambiguate speech-face associations. A simple scene of a panel discussion where all the speakers are visible for all time is an example where speakers' identity co-occurrences do not have adequate information to establish speech-face correspondences. The speakers' identity-based methods also struggle when speakers are present off-screen. However, the audio-visual activity can help overcome these limitations.  

In this work, we propose to use two sources of information, i) visual-activity-based information and ii) speakers' identity co-occurrence information, in collaboration with each other, for active speaker detection in entertainment media videos. We built this work on top of our previous work~\cite{sharma2022unsupervised}, where we proposed an unsupervised strategy for ASD using the speakers' identity co-occurrences across modalities. To gather audio-visual activity information, we use TalkNet~\cite{tao2021someone}, which shows state-of-the-art performance on the benchmark AVA-active speaker dataset~\cite{roth2020ava} for ASD. We propose a novel framework that acquires assistance from the audio-visual activity to associate speakers' identities across the audio and visual modalities for active speaker detection. We show that the acquired av-activity assistance help overcome the limitations of the speakers' identity systems. Furthermore, we show that a simple late-fusion of the scores from two methods enhances ASD performance. We evaluate the proposed approach on three datasets: the AVA-active speaker dataset~\cite{roth2020ava}, Friends, and TBBT (part of the Visual Person Clustering Dataset~\cite{brown2021face}) consisting of movies and videos from TV shows.
The contributions of this work are as follows:

\begin{enumerate}
    \item  We perform an error analysis for two ASD systems: i) audio-visual activity-based, and ii) speakers' identity-based. We show that a significant fraction of errors is exclusive to each system, highlighting the complementary nature of the two sources of information; thus supporting the need to integrate the two approaches for ASD.
    \item We propose a novel unsupervised framework to guide the speakers' identity-based speech-face associations using audio-visual activity information. The framework is generalized to ASD predictions from any external source.
    \item We propose a late fusion of the posterior scores from the audio-visual activity system and the improved speakers' cross-modal identity association system and show an enhanced ASD performance on entertainment media videos: AVA active speaker dataset (movies) and VPCD (TV shows).
\end{enumerate}

%% file: methods.tex
\section{METHODOLOGY}
\label{sec:methods}
\subsection{Problem Setup}
We use the contiguous speech segments belonging to one speaker as the fundamental speech unit for our analysis and call them speaker-homogeneous speech segments. We denote the set of all the speaker-homogeneous speech segments as $S_\text{all} \equiv \{s_1, \dots s_n, \dots s_N \}$, where $N$ is the total number of such segments. From the video frames, we use the face tracks (contiguous instances of faces belonging to one person) as the fundamental unit. For each speaker-homogeneous speech segment, $s_n$, we collect the set of face tracks that temporally overlap with the $s_n$ and denote them as $F_n \equiv \{f_1, \dots f_k,\dots f_{K_n}\}$. 

If $f_k \in F_n$ is the active speaker's face for the underlying speech segment, $s_n$, we denote it as $(s_n \longleftrightarrow f_k)$. For some speech segments, there may be no temporally overlapping face tracks. In the context of active speaker face detection, such speech segments are irrelevant; thus, we remove the speech segments where $F_n = \phi$ from the set, $S_\text{all}$. 
We only use the speech segments with at least one face track concurrently appearing with the speech segment in the video frames. 
These speech segments included cases when the active speaker's face is off-screen, even though some face tracks (none belonging to the speaker) are visible in the video frames. 

We set up the task of active speaker detection as associating a face track, $a_n$, if there exists one,  for each speaker-homogeneous speech segment, $s_n$, from the set of temporally overlapping face tracks, $F_n$, such that $a_n$ is the source face for the underlying speech segment. 
\begin{align}
    A \equiv & \{ a_n \mid a_n \in F_n \text{ and } (s_n \longleftrightarrow a_n)\} & \forall s_n \in S_\text{all}
\end{align}

\begin{figure*}
    \centering
    \includegraphics[width=0.9\textwidth,keepaspectratio]{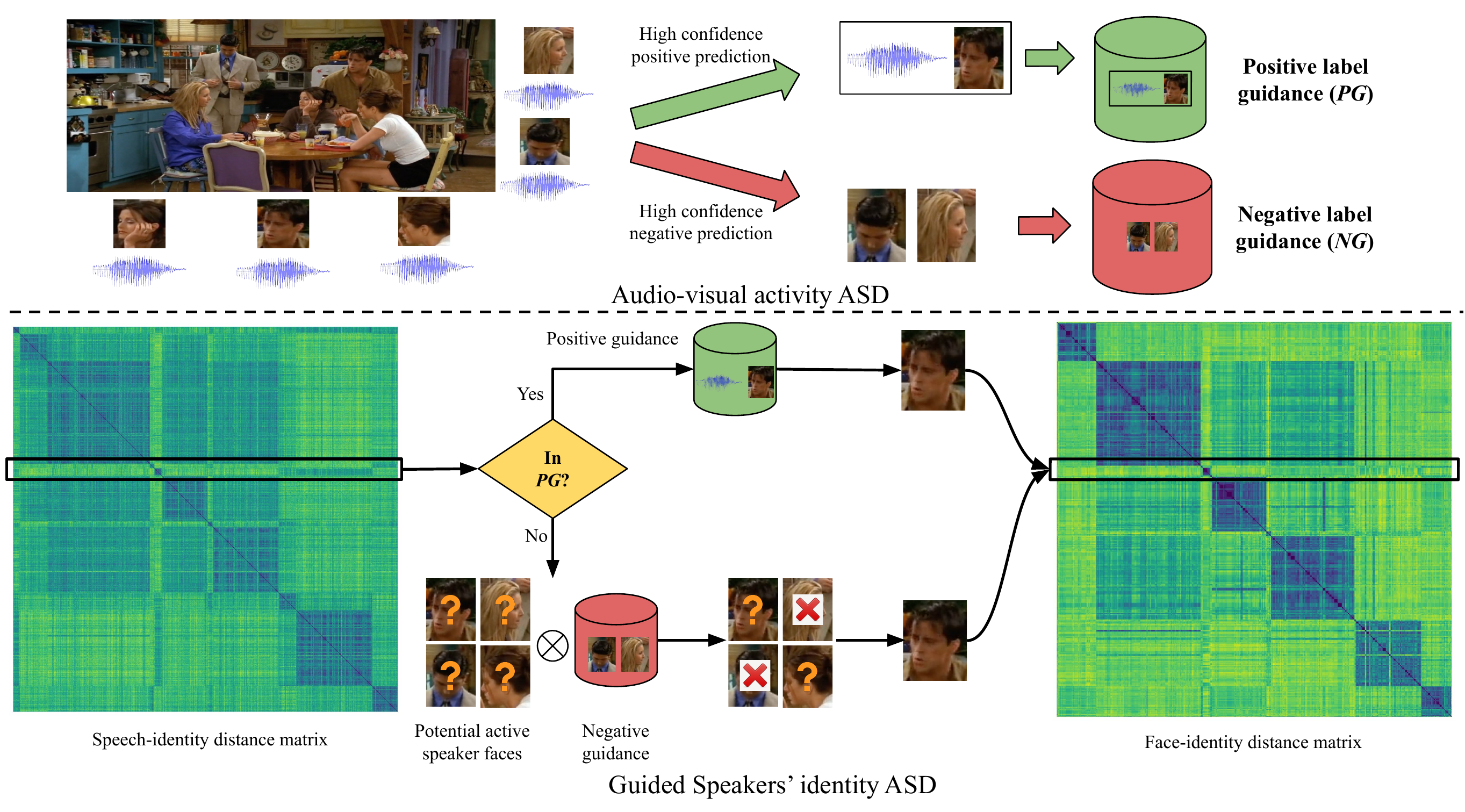}
    \caption{Overview of the audio-visual activity guided speaker identity association across modalities, \emph{GSCMIA}. a) Construction of positive and negative guides from audio-visual activity. b) Guiding \emph{SCMIA} using positive and negative guides.}
    \label{fig:overview}
\end{figure*}

\subsection{Speakers' cross-modal identity association (SCMIA)}
We use the unsupervised framework proposed in our previous work~\cite{sharma2022unsupervised}, which utilizes the speakers' cross-modal identity association for ASD. We initialize the set of active speaker faces, $A$, by randomly selecting a face track as the active speaker face for each speech segment ($s_n$) from the collection of temporally overlapping face tracks ($F_n$). To capture the identity information, we represent the speaker-homogeneous speech segments, $s_n$, using speaker recognition embeddings obtained using a ResNet~\cite{resnet} pretrained on VoxCeleb2~\cite{voxceleb2} with the angular objective function~\cite{indefense}. Similarly, we represent the faces, $f_k$, using the face recognition embeddings obtained using SENet-50~\cite{senet}, pretrained on the MS-Celeb-1M dataset~\cite{guo2016ms}. 

We construct a speech-identity distance matrix, $SD$, using all the speaker-homogeneous speech segments $s_n \in S_\text{all}$. We compute the distance between two speech segments, $s_i$ and $s_j$, using the cosine distance measure between their corresponding speaker recognition embeddings, as shown in Eqn.~\ref{eq:disatance_matrix}. The constructed matrix, $SD$, represents each speech segment in terms of its identity-based distance from all other speech segments, thus capturing the temporal identity structure existing through the video. 

Using the corresponding active speaker faces, we construct a face-identity distance matrix, $FD$, similar to $SD$, by representing each active speaker face in terms of its distance from all other active speaker faces. We compute the distance between two active speaker faces using the cosine distance between their face recognition embeddings, as shown in Eqn.~\ref{eq:disatance_matrix}. The matrix $FD$ again captures the temporal identity structure but utilizes the active speaker faces. Since the speech and the corresponding active speaker's face identify the same person, the speaker identity structure captured by the two matrices must show a high resemblance. 

\begin{align}
\label{eq:disatance_matrix}
    SD[i,j] &= \frac{s_{i} \cdot s_{j}}{\|s_{i}\| \|s_{j}\|} &
    FD[i,j] &= \frac{a_{i} \cdot a_{j}}{\|a_{i}\|\|a_{j}\|}
\end{align}

We compute the resemblance between the two distance matrices, $SD$ and $FD$, using Pearson's correlation, computed row-wise, as shown in Eqn.~\ref{eq:optimization}. We establish the speech-face correspondence by iteratively selecting an active speaker face, $a_n$, for each speech segment, $s_n$, from the set of temporally overlapping faces, $F_n$, such that the resemblance between the $SD$ and $FD$, computed using the Eqn.~\ref{eq:optimization}, is maximized. We update the set of active speaker faces iteratively until the objective function, $O(FD)$, converges.

\begin{equation}
\label{eq:optimization}
  \begin{split}
      O(FD) =\\
      \frac{1}{N}\sum_i&\frac{\sum (SD[i] - \overline{SD[i]})(FD[i] - \overline{FD[i]})}{\sqrt{\sum(SD[i] - \overline{SD[i]})^2\sum(FD[i] - \overline{FD[i]})^2}}
  \end{split}
\end{equation}
\begin{equation}
     A \equiv \argmax_{\{ a_n \mid a_n = f_{k} \text{ and } f_k \in F_n\}}Corr(FD)
\end{equation}

Post convergence, this framework provides an active speaker face track for each speaker-homogeneous speech segment. It even provides an active speaker face for the speech segments where speakers are present off-screen, which adds to the false positives of the system. Moreover, this framework is limited to the videos having enough information to disambiguate characters in the video. One of the failure cases is a single-scene video where all speakers are visible in frames at all times. In such a scenario, the system will have no way to associate the speech and faces correctly. For brevity, we will refer to this system, using speakers' cross-modal identity associations, as \emph{SCMIA}. 

\subsection{Audio-visual activity guidance (\emph{GSCMIA})}
\label{subsec:avg}
To address the limitations of the \emph{SCMIA}, we propose incorporating the audio-visual activity information, which can complement the speaker's identity information. To gather the audio-visual activity information, we use a state-of-the-art audio-visual model, TalkNet~\cite{tao2021someone}. TalkNet observes a face track and the concurrent audio waveform and models the short-term and long-term temporal context to provide face-box-wise active speaker predictions. Using the obtained face box scores we compute an active speaker score, $\hat{P_a}(f_k)$ for each face track, $f_k$, as the mean score of the constituent face boxes.

We use the obtained TalkNet predictions to inject the disambiguating information into the \emph{SCMIA}. To minimize the amount of noise introduced by the audio-visual TalkNet, we utilize only the highly confident predictions, whether positive or negative predictions. We use the predictions in two ways:
\newline
\textit{i) Positive-label guidance:} We intend to collect the speech segments for which the audio-visual activity predictions provide a highly confident speech-face association. For each speech segment $s_n$, we select the face track from the set of temporally overlapping face tracks, $t_n \in F_n$, that have the maximum AV-activity-based active speaker posteriors, $\hat{P}_a(t_n)$. The speech segments with selected face tracks, $t_n$, displaying a highly confident positive posterior are collected; we call it the positive guidance set $(PG)$, denoted in Eqn.~\ref{eq:PG}. For all these speech segments, we use the AV-activity predictions as the ground truth and assign the selected face track, $t_n$, as the active speaker face (Eqn.~6).  We keep these speech-face assignments unaltered while iteratively maximizing the $O(FD)$ in Eqn.~\ref{eq:optimization} \& 4. Explicitly assigning the selected face track to the speech segment replicates the case when only the speaker's face is visible in the video frames. Such cases are especially disambiguating as it provides a trivial signal for the speech-face association.
\begin{align}
    \label{eq:PG}
     t_n = &\argmax_{f_k \in F_n}\hat{P_a}(f_k) & PG &\equiv \{s_n \mid \hat{P_a}(t_n) > \tau_p\} \\
     & (s_n \longleftrightarrow t_n)  &  \forall &  s_n \in PG
\end{align}
\textit{ii) Negative-label guidance:} We intend to take advantage of the highly confident negative predictions by the AV-activity system. For each speech segment, $s_n$, we collect the face tracks having highly confident negative av-activity predictions, and call them a negative guidance set $(NG)$, denoted in Eqn.~\ref{eq:NG}. Then we remove the face tracks in $NG_n$ from the candidate active speaker faces, $F_n$, as they correspond to non-speaking faces. It makes the set, $F_n$ smaller and makes it easier to find the active speaker face for the underlying speech segment. It also addresses the speech segments with off-screen speakers, removing all the non-speaking faces, thus leaving no faces in the candidate active speaker faces, $F_n$. We call this audio-visual activity guided speaker identity association as \emph{GSCMIA}. The overview of \emph{GSCMIA} is shown in Figure~\ref{fig:overview} and pseudo-code in Algorithm~\ref{algo:optimzation}.
\begin{align}
    \label{eq:NG}
    NG_n &= \{f_k \mid f_k \in F_n, \hat{P_a}(f_k) < \tau_n\} & F_n &= F_n - NG_n
\end{align}

\begin{algorithm}[tb]
\label{algo:optimzation}
$\text{Positive guidance:} \equiv PG$ \tcp*{using eq:~\ref{eq:PG}}
$a_n = t_n \forall s_n \in PG$\;
$\text{Negative guidance:} \equiv NG_n$ \tcp*{using eq:~\ref{eq:NG}}
$F_n = F_n - NG_n \forall s_n \in S_\text{all}$\;
$S_\text{all} = S_\text{all} - \{s_n \mid F_n = \phi \}$\;
$\text{Randomly initialize:} A \equiv \{a_n \mid  a_{n} \in F_{n}\} \forall s_n \in S_\text{all} $\;
Compute $SD$ and $FD$ \tcp*{using eq:~\eqref{eq:disatance_matrix}}
$objective \gets Corr(FD)$ \tcp*{using eq:~\eqref{eq:optimization}} 
\While{$objective$ increases}{
    \For{each $a_i \in A$}{
    \If{$s_i \in PG$}{
    $continue$ \tcp*{remains unaltered}
    }
    $a_{i} = \argmax\limits_{f_k \in F_{n}} Corr(FD)$ \tcp*{$i^{th}$ row}
    $\text{Update } FD$\;
    $\text{Update } A$\;
    }
    $ \text{$objective$} \gets Corr(FD)$\;
  }
  \caption{GSCMIA}
\end{algorithm}

%% file: expt.tex
\section{EXPERIMENTS AND IMPLEMENTATION DETAILS}
\subsection{Implementation details}
We start with extracting the active voice regions for a video under consideration using an open-source tool, \emph{pyannote}~\cite{pyannote}. We then segment the obtained voiced parts into approximated speaker-homogeneous speech segments, $s_n$, by partitioning at the shot boundaries~\cite{pardo2021moviecuts} and keeping the maximum duration to be 1sec each~\cite{sharma2022unsupervised}. We gather the set of temporally overlapping face tracks, $F_n$, for all the obtained speech segments, using the RetinaFace~\cite{deng2020retinaface} face detector, and SORT~\cite{Bewley2016_sort} tracker. As a trade-off between the computational time and the context length~\cite{sharma2022unsupervised}, we partition the set of speech segments, $S_\text{all}$, to contain $L$ segments each. Next, we compute each partition's speech-identity and face-identity distance matrix and solve the speech-face association by separately maximizing each partition's objective function $O(FD)$. For \emph{SCMIA} we use $L=500$. 

For \emph{GSCMIA} we employ TalkNet~\cite{tao2021someone} to gather AV-activity ASD predictions. We construct the positive guidance set of speech segments, $PG$, using $\tau_p > 0.9$ (Eqn.~\ref{eq:PG}) for AVA dataset and $\tau_p > 0.5$ for VPCD videos, where TalkNet prediction scores lie in the range $[0,1]$. While constructing the negative guidance set of face tracks, $NG_n$, we use $\tau_n < 0.2$ (Eqn.~\ref{eq:NG}) for both the datasets. These thresholds are derived using a subset of training sets for both the datasets. In contrast to ~\emph{SCMIA}, we use $L=50$ for \emph{GSCMIA}.

For evaluation purposes we use two benchmark datasets: i) the AVA-active speaker dataset~\cite{roth2020ava} and ii) the Visual person clustering dataset~\cite{brown2021face}. The AVA-active speaker dataset consists of 160 international movies, publicly available, and face-boxwise annotations for 15 min duration of each film. We report performance for the publicly available and widely used validation split of the AVA-active speaker dataset consisting of 33 movies. To incorporate diversity, we use a part of the VPCD consisting of active speaker annotations and character identities for several Hollywood TV shows and movies. We observed that the available annotations in VPCD are exhaustive only for TV shows: Friends and TBBT, primarily due to the limited number of characters in the videos. So we restrict our evaluation to TV shows: 25 episodes of Friends from season 3 and 6 episodes of TBBT from season 1. We use the widely used mean average precision (mAP) metric to report performances.

\subsection{Performance Evaluation}
In Table~\ref{tab:mAP}, we report the performance of various strategies described in Section~\ref{sec:methods} for the three datasets. We present the performance of \emph{SCMIA} and TalkNet~\cite{tao2021someone}, computed as an average over mAP of the constituting videos for each dataset. We compute the mAP using the official tool provided by AVA-active speaker. We observe that TalkNet performs significantly better than \emph{SCMIA} for the AVA dataset. At the same time, the performance is comparable for both strategies in the case of the TV shows: Friends and TBBT. A reason for such observed behavior can be the fully-supervised setup of TalkNet, trained on AVA training set. Table~\ref{tab:mAP} also shows the performance of \emph{GSCMIA}, following Algorithm~\ref{algo:optimzation}. The av-activity guidance enhances the performance of \emph{SCMIA} across all datasets. We note that the performance enhancement for the AVA dataset is more significant than for others.

Since the sources of information for audio-visual activity in the case of TalkNet and \emph{SCMIA} is independent, we expect them to have some exclusive components. Here we investigate the validity of this hypothesis by combining the two methods at the score level. We score each face box as the weighted aggregate of the individual scores from the two methods: i) \emph{GSCMIA} and ii) TalkNet, in a ratio of $\alpha:(1-\alpha)$.
We report the obtained mAP for the three datasets in Table~\ref{tab:mAP}. We observe that a simple late fusion of the two methods improved the performance for all three datasets, showing performance better than the individual performances of the involved methods.
\begin{table}[]
\centering
\caption{Performance of various strategies for ASD when guided with Talknet~\cite{talknet}. The reporting metric is mean average precision (\%).}
\label{tab:mAP}
\resizebox{0.45\textwidth}{!}{%
\begin{tabular}{@{}c|cccc@{}}
\toprule
Datasets & \emph{SCMIA} & TalkNet & \emph{GSCMIA} & Late-fusion \\ \midrule
\begin{tabular}[c]{@{}c@{}}AVA\\ active speaker\end{tabular}    & 68.84 & 91.96 & 80.9  & \textbf{92.86} \\ \midrule
\begin{tabular}[c]{@{}c@{}}Friends\\ (25 episodes)\end{tabular} & 81.45 & 85.42 & 84.13 & \textbf{87.54}                         \\ \midrule
\begin{tabular}[c]{@{}c@{}}TBBT\\ (6 episodes)\end{tabular}     & 85.74 & 83.84 & 86.39 & \textbf{87.19}                     \\ \bottomrule
\end{tabular}%
}
\end{table}

We further evaluate the \emph{GSCMIA} guided with another av-activity model: syncnet~\cite{chung2016out}. Syncnet is trained for audio-visual synchronization in a self-supervised framework, unlike fully-supervised TalkNet. In Table~\ref{tab:syncnet}, we show the performance of Syncnet and \emph{GSCMIA} assisted with Syncnet, for the three datasets. We note the weaker performance of Syncnet compared to TalkNet. We observe that the assistance from Syncnet enhances the performance for \emph{GSCMIA}, although marginally for TV shows, attributing to the weaker performance of SyncNet. The two methods' late fusion of posterior scores performs improves the overall performance. The observed trends in performance enhancement with SyncNet and TalkNet guidance are consistent.

\begin{table}[]
\centering
\caption{Performance of various strategies for ASD when guided with Syncnet~\cite{chung2016out}. The reporting metric is mean average precision (\%).}
\label{tab:syncnet}
\resizebox{0.45\textwidth}{!}{%
\begin{tabular}{@{}c|cccc@{}}
\toprule
Datasets & \emph{SCMIA} & SyncNet & \emph{GSCMIA} & Late-fusion \\ \midrule
\begin{tabular}[c]{@{}c@{}}AVA\\ active speaker\end{tabular}    & 68.84 & 58.70 & 71.74  & \textbf{73.74} \\ \midrule
\begin{tabular}[c]{@{}c@{}}Friends\\ (25 episodes)\end{tabular} & 81.45 & 77.14 & 81.98 & \textbf{83.90}                         \\ \midrule
\begin{tabular}[c]{@{}c@{}}TBBT\\ (6 episodes)\end{tabular}     & 85.74 & 80.00 & 85.87 & \textbf{86.14}                     \\ \bottomrule
\end{tabular}%
}
\end{table}
\subsection{Ablation studies}
\subsubsection{Error Analysis}
In this work, we considered two sources of information concerning active speaker detection in videos: i) the audio-visual activity modeling and ii) the speaker identity co-occurrences across modalities. The fundamental hypothesis underlying this work is that the information captured by the two sources may have complementary components. Here we investigate this further by comparing the predictions of the two systems. 


\begin{figure}[]
    \centering
    \includegraphics[width=0.5\textwidth,keepaspectratio]{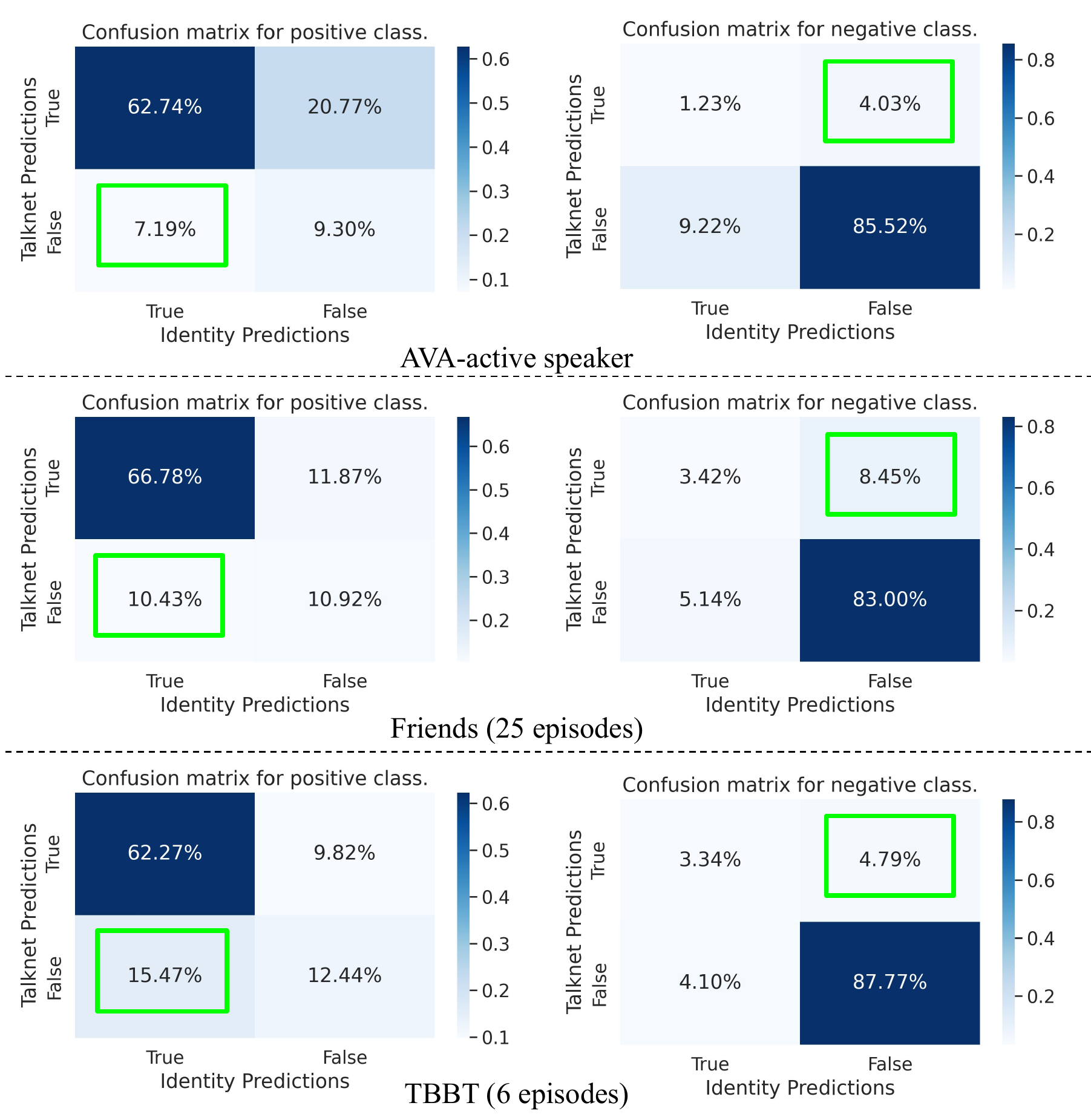}
    \caption{Comparison of \emph{SCMIA} and TalkNet~\cite{tao2021someone} perdictions for positive (top row) and negative samples (bottom row).} 
    \label{fig:error_analysis}
\end{figure}
We generate face box-wise scores using i) \emph{SCMIA} and ii) TalkNet for the three datasets. We then generate the corresponding predictions by thresholding the obtained scores such that the systems' precision and recall are nearly equivalent. We divide the samples into two groups: i) positive samples and ii) negative samples, using ground truth. For both groups, we construct the confusion matrices, comparing the predictions of the two systems. In Figure~\ref{fig:error_analysis} we show the confusion matrices for the positive and negative samples for three datasets (AVA-active speaker, Friends, and TBBT). 

The off-diagonal cells in confusion matrices shown in Figure~\ref{fig:error_analysis} represent the exclusive correctness of one of the methods. The cells marked in green shows the fraction of samples where the \emph{SCMIA} is exclusively correct. We note that the fraction of samples where one method is solely accurate is significant, especially for the positive samples. From Table~\ref{tab:mAP}, we note that the performance of TalkNet on AVA is significantly higher than the \emph{SCMIA}, precisely 92.0 mAP score, and has marginal room for further improvement. Even then, the fraction of exclusive correctness of the \emph{SCMIA} is significant. 
This verifies the hypothesis that the information comprising the audio-visual activity and \emph{SCMIA} have some exclusive components, which motivates us to explore the use of two sources of information in a complementary fashion towards enhancing ASD performance. 


\subsubsection{Role of Positive guidance $(PG)$ in \emph{GSCMIA}}
The Positive guides intend to provide explicit disambiguating information to the \emph{SCMIA} by acquiring high-confidence positive class (speaking) predictions from the audio-visual activity information (TalkNet). The speech segments in $PG$ and their corresponding active-speaker face tracks readily provide the speech-face associations to the \emph{SCMIA}. To quantify the role of $PG$, we report, in Table~\ref{tab:pg}, the fraction of all speech segments which provide positive guidance: $\frac{|PG|}{|S_\text{all}|}$, where $|.|$ denotes the cardinality of a set. We also report the accuracy of the positive guides toward predicting the positive class (speaking). We observe that the accuracy of the PG is high for all the datasets. This validates the usage of the $PG$ as the ground truth, adding a marginal amount of noise in the form of a small number of false positives. We note that the $PG$ fraction is larger and accuracy is smaller for TV shows compared to AVA. This is due to the lower value of $\tau_p$ for VPCD.


For some of the speech segments in $PG$, speech-face associations are correctly deduced even by the \emph{SCMIA} (no-guidance). A simple example is a case when only the speaker's face is visible in the frames. Effectively the speech segments in $PG$, for which the \emph{SCMIA} can establish the same speech-face association, do not help other than reducing the computational load. Removing such speech segments from the $PG$, we report the fraction of the $S_\text{all}$ that presents effectively helpful information in Table~\ref{tab:pg} as Effective positive guides.  We observe that a much smaller fraction of $S_\text{all}$ constitutes the effective positive guides.
We also tabulate the accuracy of the effective positive guides and observe that they are less accurate than the bigger set $PG$. The observed smaller accuracy is notable for TV shows, attributed to the lower value of $\tau_p$ for VPCD. 
\begin{table}[]
\centering
\caption{Constituents of positive and effective positive guides and their exactness. All values are shown in \%.}
\label{tab:pg}
\resizebox{0.45\textwidth}{!}{%
\begin{tabular}{@{}c|cc|cc@{}}
\toprule
    \multirow{2}{*}{Videos}               & \multicolumn{2}{c|}{Positive guides (PG)} & \multicolumn{2}{c}{Effective PG} \\ \cmidrule(l){2-5} 
             & Fraction         & Accuracy         & Fraction              & Accuracy             \\ \midrule
AVA & 24.65            & 91.67             & 4.41                 & 82.03                 \\
Friends            & 73.42            & 79.69             & 12.09                  & 47.73                 \\
TBBT               & 66.59            & 83.13             & 7.81                  & 65.82                 \\ \bottomrule
\end{tabular}%
}
\end{table}

\subsubsection{Positive guides at work}
Here we display the advantage of the positive guidance from the audio-visual activity to solve the speech-face association using \emph{GSCMIA}. We consider an adverse case where the speaker identity-based method fails due to insufficient disambiguating information. One such scenario is where all the speaker faces are always visible in frames; a panel discussion is a simple example. The Columbia dataset, consisting of an 85 min long panel discussion video, is the closest, and it consists of active speaker annotations for 35 min duration of the video comprising six speakers.

We evaluate the performance of the \emph{SCMIA} (no guidance), on the Columbia dataset and report the widely used speaker-wise weighted F1 score in Table~\ref{tab:columbia}. We also report the performance of several state-of-the-art methods for comparison. We observe that the performance is terrible, particularly for three speakers: Lieb, Long, and Sick, while comparable to other methods for the remaining speakers. By visually inspecting the system's active speaker predictions, we noted that the system incorrectly associates Long's speech with Sick's face and Lieb's with Long's face. The relevant fragments of the video output are present in the supplementary material. These erroneous associations can be attributed to the fact that Lieb and Sick always appear together in the video, providing the system with insufficient information to resolve the speech-face association. 
\begin{table}[tb]
\centering
\caption{Comparison of the speaker-wise weighted F1 scores for all the speakers in Columbia dataset.}
\label{tab:columbia}
\resizebox{0.45\textwidth}{!}{%
\begin{tabular}{@{}ccccccc|c@{}}
\toprule
Methods             &Abbas  & Bell & Boll  & Lieb & Long & Sick & Avg   \\ \midrule
Chakravarty et al.~\cite{chakravarty2016cross}  &-      & 82.9 & 65.8  & 73.6 & 86.9 & 81.8 & 78.2  \\
Shahid et al~\cite{shahid_columbia}          &-      & 89.2 & 88.8  & 85.8 & 81.4 & 86   & 86.2  \\
Chung et al.~\cite{chung2016out}             &-      & 93.7 & 83.4  & 86.8 & 97.7 & 86.1 & 89.5  \\
Afouras et al.~\cite{afouras2020self}     &-      & 92.6 & 82.4  & 88.7 & 94.4 & 95.9 & 90.8  \\
S-VVAD~\cite{svvad}           &-      & 92.4 & 97.2  & 92.3 & 95.5 & 92.5 & 94    \\
RealVAD~\cite{realvad}           &-      & 92   & 98.9  & 94.1 & 89.1 & 92.8 & 93.4  \\
TalkNet ~\cite{talknet}           &-      & 97.1 & 90.0    & 99.1 & 96.6 & 98.1 & 96.2  \\ \midrule
\emph{SCMIA}           &97.0  & 95.0 & 91.0 & 65.3  & 0.1 & 34.6  & 65.0 \\
\emph{GSCMIA} &97.3  & 96.3 & 89.4  & 98.7 & 98.7 & 96.8 & \textbf{96.2}  \\ \bottomrule
\end{tabular}%
}
\end{table}

We inject the explicit disambiguating information in form of positive guides from TalkNet. We use just the positive guides, no negative guides, while employing Algorithm~\ref{algo:optimzation} to quantify the impact of the positive guides. We report the performance of the positively guided \emph{GSCMIA} for all the speakers in Table~\ref{tab:columbia}. We observe that the guided system restores the earlier incorrect speech-face associations and performs at par with other methods for all the speakers. The relevant video fragments with enhanced predictions are present in the supplementary material. The improved performance highlights the importance of positive guides in providing the required disambiguating information to establish accurate speech-face associations. 

\subsubsection{Role of Negative guides ($NG_n$)}
The visual information modeling the activity in the lip region and its coordination with the underlying audio waveform can reliably predict which of the visible faces are not speaking. The negative guides are introduced to incorporate the information concerning the non-speaking faces to help the speech-face association using the speaker identity contextual information. The face tracks classified by the TalkNet as not speaking with high enough confidence are used as ground truth. We remove these faces from the set of candidate active speaker faces while establishing speech-face association using the \emph{GSCMIA}.

Although not explicitly as in the case of positive guides, the negative guides also introduce disambiguating information by eliminating some of the candidate active speaker faces. An example can be a scenario with two faces visible in the frames, and one of them is eliminated by the negative guides classifying as not speaking: establishing a direct speech-face correspondence. In Table~\ref{tab:ng}, we quantify the fraction of all face tracks that constitute the negative guides: $\frac{\sum_n|NG_n|}{\sum_n|F_n|}$, along with their accuracy predicting the negative class (not speaking). We observe high accuracy values for all datasets, attributed to selecting only high-confidence predictions from TalkNet. We also note that the fraction of face tracks that negative guides constitutes is much higher for AVA than others. The reason can be that TalkNet performs better for AVA than others, primarily due to training on the AVA train set. 
\begin{table}[]
\centering
\caption{Constituents of negative and effective negative guides and their exactness. All values are shown in \%.}
\label{tab:ng}
\resizebox{0.45\textwidth}{!}{%
\begin{tabular}{@{}c|cc|cc@{}}
\toprule
     \multirow{2}{*}{Videos}               & \multicolumn{2}{c|}{Negative guides (NG)} & \multicolumn{2}{c}{Effective NG} \\ \cmidrule(l){2-5} 
           & Fraction            & Accuracy           & Fraction       & Accuracy       \\ \midrule
AVA & 46.69               & 96.06               & 20.54          & 89.57           \\
Friends            & 22.00               & 98.43               & 2.98           & 86.72           \\
TBBT               & 17.87               & 93.87               & 3.61           & 56.59           \\ \bottomrule
\end{tabular}%
}
\end{table}

As in the case of $PG$, only some of the face tracks in negative guides provide new information to the \emph{GSCMIA}. Even with no guidance, the \emph{SCMIA} can correctly deduce some of the face tracks in negative guides as not speaking. Discarding such face tracks, we report the remaining fraction of all face tracks in Table~\ref{tab:ng} as effective negative guides, along with their precision for negative class. We observe that only a small fraction of face tracks in $NG_n$ effectively add value to \emph{GSCMIA} for TV shows while a notable fraction for the AVA dataset. Surprisingly, we note that the accuracy of the effective negative guides is low for TBBT, adding noisy information. The noisy guides lead to only marginal increase in the overall performance with the av-guidance, \emph{GSCMIA}, for TBBT compared to others, as shown in Table~\ref{tab:mAP}

\subsubsection{Effects on offscreen speakers}
One of the limitations of \emph{SCMIA} is the inability to explicitly address the speech segments where faces are visible in the frames, but the speaker is off-screen. However, audio-visual activity methods can handle such cases reliably. The negative guides are intended to fuse this ability of the AV-activity information with the \emph{SCMIA}. Here we explore the effect of AV-activity guidance on the speech segments with off-screen speakers.

We tabulate the fraction of speech segments with off-screen speakers in Table~\ref{tab:offscreen} for all three datasets. We observe that the AVA has much higher fraction of speech segments with off-screen speakers than others. It is due to the nature of videos in AVA which are more in the wild compared to structured TV shows where the video primarily captures the speaker. The higher fraction of off-screen speakers in AVA is one of the reasons for the lower performance of the \emph{SCMIA} for AVA compared to others (Table~\ref{tab:mAP}). We compare the errors of \emph{SCMIA} and \emph{GSCMIA}, particular to speech segments with off-screen speakers. We report the fraction of speech segments with off-screen speakers for which the systems incorrectly assign an active speaker face in Table~\ref{tab:offscreen}.
\begin{table}[]
\centering
\caption{Effect of using the Negative guides: performance enhancement for speech segments with off-screen speakers.}
\label{tab:offscreen}
\resizebox{0.4\textwidth}{!}{%
\begin{tabular}{@{}c|ccc@{}}
\toprule
Videos  & off-screen (\%) & \emph{SCMIA} & \emph{GSCMIA} \\ \midrule
AVA     & 33.30           & 16.67       & 9.51                 \\
Friends & 13.31           & 19.64       & 17.28                \\
TBBT    & 12.21           & 10.32       & 10.02                \\ \bottomrule
\end{tabular}%
}
\end{table}

We observe that the errors in the form of false positives by assigning active speaker faces to speech segments with off-screen speakers reduce consistently for all datasets when guided with av-activity. We point out that the improvement in the AVA dataset is greater than TV shows. The larger fraction of effective negative guides for AVA dataset, observed in Table~\ref{tab:ng}, which in turn are due to large fraction of off-screen speakers, is the reason for the observed substantial improvement.

\subsubsection{Effects on required temporal context}
\begin{figure}[]
    \centering
    \includegraphics[width=0.5\textwidth,keepaspectratio]{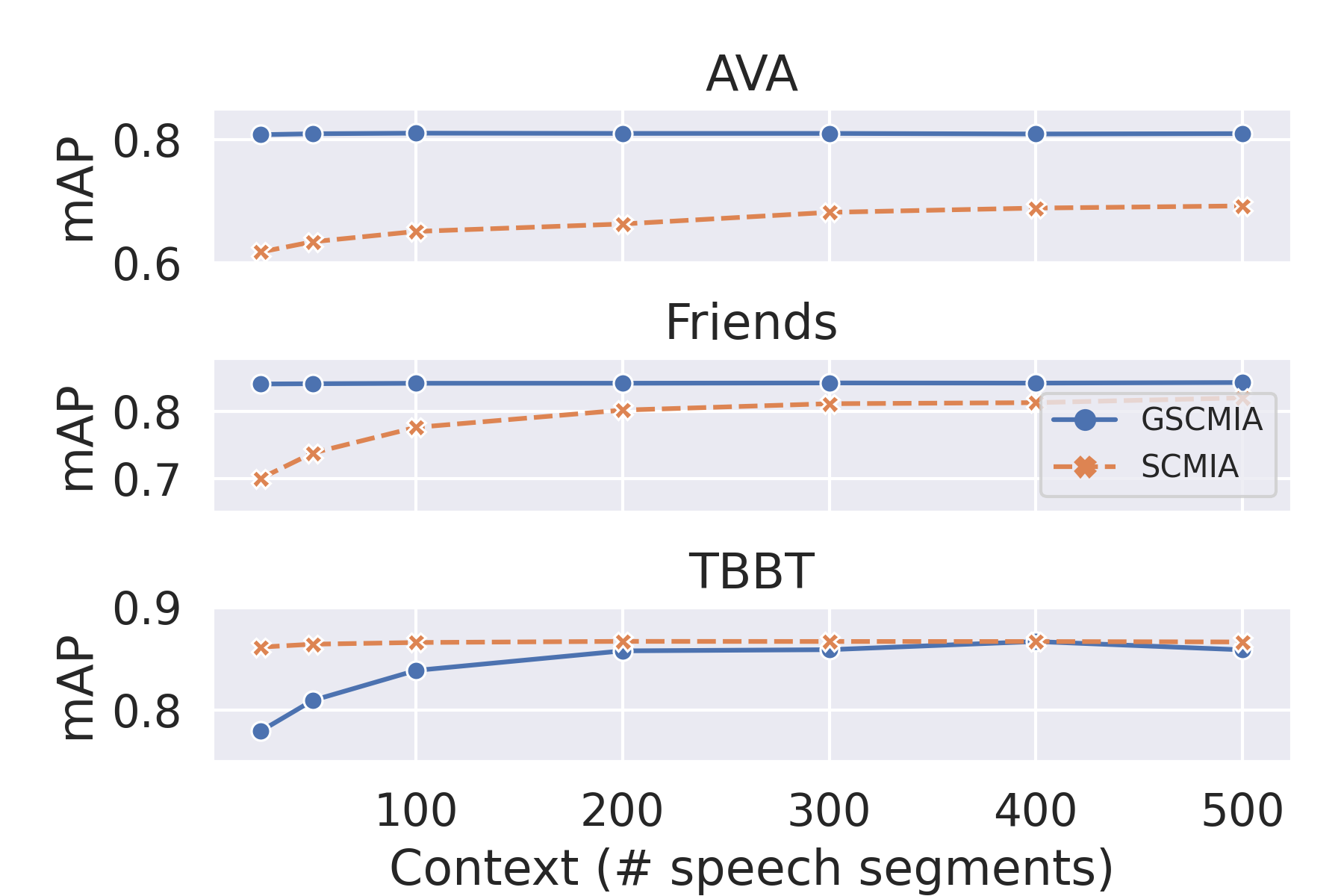}
    \caption{Variation in performance, reported in mAP, of \emph{SCMIA} and \emph{GSCMIA} with the context length $(L)$}
    \label{fig:context}
\end{figure}
The \emph{SCMIA} studies the contextual information concerning the speakers' identity appearing in the speech and faces and establishes the speech-face associations. By formulation, it relies on the observed speakers' identities in temporal context to have enough disambiguating information to solve the required speech-face associations. Using all the available speech segments, i.e., the entire video length, is ideal. However, as described in ~\cite{sharma2022unsupervised}, the complexity of finding the speech-face association, using the Algorithm~\ref{algo:optimzation}, is a quadratic function of the context length ($O(L^2)$); thus, a lower value of $L$ is desirable. The value of $L$ acts as a tradeoff between the system's performance and time complexity.

As mentioned in the implementation details, we use $L=500$ for \emph{SCMIA}. The system observes the context of identities from speech and faces appearing during the 500 speech segments and assumes that it consists of enough disambiguating information. At the same time, we use a context of only 50 segments $(L=50)$ while employing \emph{GSCMIA}. Here we explore the role of required context of speech segments to solve speech-face associations.

In Figure~\ref{fig:context}, we show the variation in the performance of each system (\emph{SCMIA} \& \emph{GSCMIA}) with the length of context ($L$). We observe a notable increase in the performance (reported in mAP) of the \emph{SCMIA} with the increase in $L$. This increased performance follows from the fact that the system observes more disambiguating information with a larger context length ($L$) and thus better solves the speech-face association. In contrast, the system's performance with av-activity guidance (\emph{GSCMIA}) is robust to the variations in the context length. It validates the hypothesis that the AV-activity guides, positive and negative, contribute to the disambiguating information making even the smaller context length enough to solve the speech-face associations. We also observe that the performance of the guided system with just 25 speech segments is significantly better than the unguided system, even with a context of 500 speech segments. Thus guiding the speaker identity-based system with AV-activity information shows a 400-times improvement in the time complexity along with the noted improvement in performance.

%% file: main.bbl
\begin{thebibliography}{10}
\providecommand{\url}[1]{#1}
\csname url@samestyle\endcsname
\providecommand{\newblock}{\relax}
\providecommand{\bibinfo}[2]{#2}
\providecommand{\BIBentrySTDinterwordspacing}{\spaceskip=0pt\relax}
\providecommand{\BIBentryALTinterwordstretchfactor}{4}
\providecommand{\BIBentryALTinterwordspacing}{\spaceskip=\fontdimen2\font plus
\BIBentryALTinterwordstretchfactor\fontdimen3\font minus
  \fontdimen4\font\relax}
\providecommand{\BIBforeignlanguage}[2]{{%
\expandafter\ifx\csname l@#1\endcsname\relax
\typeout{** WARNING: IEEEtran.bst: No hyphenation pattern has been}%
\typeout{** loaded for the language `#1'. Using the pattern for}%
\typeout{** the default language instead.}%
\else
\language=\csname l@#1\endcsname
\fi
#2}}
\providecommand{\BIBdecl}{\relax}
\BIBdecl

\bibitem{somandepalli2021computational}
K.~Somandepalli, T.~Guha, V.~R. Martinez, N.~Kumar, H.~Adam, and S.~Narayanan,
  ``Computational media intelligence: Human-centered machine analysis of
  media,'' \emph{Proceedings of the IEEE}, vol. 109, no.~5, 2021.

\bibitem{sharma2022using}
R.~Sharma and S.~Narayanan, ``Using active speaker faces for diarization in tv
  shows,'' 2022.

\bibitem{intel_diariz}
\BIBentryALTinterwordspacing
K.~Min, ``Intel labs at ego4d challenge 2022: A better baseline for
  audio-visual diarization,'' 2022. [Online]. Available:
  \url{https://arxiv.org/abs/2210.07764}
\BIBentrySTDinterwordspacing

\bibitem{asr_google}
O.~Braga and O.~Siohan, ``A closer look at audio-visual multi-person speech
  recognition and active speaker selection,'' in \emph{ICASSP 2021 - 2021 IEEE
  International Conference on Acoustics, Speech and Signal Processing
  (ICASSP)}, 2021, pp. 6863--6867.

\bibitem{sharma2022audio}
R.~Sharma and S.~Narayanan, ``Audio visual character profiles for detecting
  background characters in entertainment media,'' \emph{arXiv preprint
  arXiv:2203.11368}, 2022.

\bibitem{deepstar}
I.~U. Haq, K.~Muhammad, A.~Ullah, and S.~W. Baik, ``Deepstar: Detecting
  starring characters in movies,'' \emph{IEEE Access}, vol.~7, pp. 9265--9272,
  2019.

\bibitem{sharma_publicspeaker}
R.~Sharma, T.~Guha, and G.~Sharma, ``Multichannel attention network for
  analyzing visual behavior in public speaking,'' in \emph{2018 IEEE Winter
  Conference on Applications of Computer Vision (WACV)}, 2018, pp. 476--484.

\bibitem{owens2018audio}
A.~Owens and A.~A. Efros, ``Audio-visual scene analysis with self-supervised
  multisensory features,'' in \emph{Proceedings of the European Conference on
  Computer Vision (ECCV)}, 2018, pp. 631--648.

\bibitem{bendris2010lip}
M.~Bendris, D.~Charlet, and G.~Chollet, ``Lip activity detection for talking
  faces classification in tv-content,'' in \emph{International conference on
  machine vision}, 2010, pp. 187--190.

\bibitem{arandjelovic2018objects}
R.~Arandjelovic and A.~Zisserman, ``Objects that sound,'' in \emph{Proceedings
  of the European conference on computer vision (ECCV)}, 2018, pp. 435--451.

\bibitem{chung2016out}
J.~S. Chung and A.~Zisserman, ``Out of time: automated lip sync in the wild,''
  in \emph{Asian conference on computer vision}.\hskip 1em plus 0.5em minus
  0.4em\relax Springer, 2016, pp. 251--263.

\bibitem{chung2019naver}
J.~S. Chung, ``Naver at activitynet challenge 2019--task b active speaker
  detection (ava),'' \emph{arXiv preprint arXiv:1906.10555}, 2019.

\bibitem{zhao2018sound}
H.~Zhao, C.~Gan, A.~Rouditchenko, C.~Vondrick, J.~McDermott, and A.~Torralba,
  ``The sound of pixels,'' in \emph{Proceedings of the European conference on
  computer vision (ECCV)}, 2018, pp. 570--586.

\bibitem{afouras2020self}
T.~Afouras, A.~Owens, J.~S. Chung, and A.~Zisserman, ``Self-supervised learning
  of audio-visual objects from video,'' in \emph{European Conference on
  Computer Vision}.\hskip 1em plus 0.5em minus 0.4em\relax Springer, 2020, pp.
  208--224.

\bibitem{sharma2020cross}
R.~Sharma, K.~Somandepalli, and S.~Narayanan, ``Cross modal video
  representations for weakly supervised active speaker localization,''
  \emph{arXiv e-prints}, pp. arXiv--2003, 2020.

\bibitem{chakravarty2015s}
P.~Chakravarty, S.~Mirzaei, T.~Tuytelaars, and H.~Van~hamme, ``Who's speaking?
  audio-supervised classification of active speakers in video,'' in
  \emph{Proceedings of the 2015 ACM on International Conference on Multimodal
  Interaction}, 2015, pp. 87--90.

\bibitem{chakravarty2016cross}
P.~Chakravarty and T.~Tuytelaars, ``Cross-modal supervision for learning active
  speaker detection in video,'' in \emph{European Conference on Computer
  Vision}.\hskip 1em plus 0.5em minus 0.4em\relax Springer, 2016, pp. 285--301.

\bibitem{sharma2019icip}
R.~Sharma, K.~Somandepalli, and S.~Narayanan, ``Toward visual voice activity
  detection for unconstrained videos,'' in \emph{2019 IEEE International
  Conference on Image Processing (ICIP)}.

\bibitem{roth2020ava}
J.~Roth, S.~Chaudhuri, O.~Klejch, R.~Marvin, A.~Gallagher, L.~Kaver,
  S.~Ramaswamy, A.~Stopczynski, C.~Schmid, Z.~Xi \emph{et~al.}, ``Ava active
  speaker: An audio-visual dataset for active speaker detection,'' in
  \emph{ICASSP 2020-2020 IEEE International Conference on Acoustics, Speech and
  Signal Processing (ICASSP)}.\hskip 1em plus 0.5em minus 0.4em\relax IEEE,
  2020, pp. 4492--4496.

\bibitem{kim2021look}
Y.~J. Kim, H.-S. Heo, S.~Choe, S.-W. Chung, Y.~Kwon, B.-J. Lee, Y.~Kwon, and
  J.~S. Chung, ``Look who's talking: Active speaker detection in the wild,''
  \emph{arXiv preprint arXiv:2108.07640}, 2021.

\bibitem{brown2021face}
A.~Brown, V.~Kalogeiton, and A.~Zisserman, ``Face, body, voice: Video
  person-clustering with multiple modalities,'' in \emph{Proceedings of the
  IEEE/CVF International Conference on Computer Vision}, 2021, pp. 3184--3194.

\bibitem{huang2020improved}
C.~Huang and K.~Koishida, ``Improved active speaker detection based on optical
  flow,'' in \emph{Proceedings of the IEEE/CVF Conference on Computer Vision
  and Pattern Recognition Workshops}, 2020, pp. 950--951.

\bibitem{alcazar2020active}
J.~L. Alc{\'a}zar, F.~Caba, L.~Mai, F.~Perazzi, J.-Y. Lee, P.~Arbel{\'a}ez, and
  B.~Ghanem, ``Active speakers in context,'' in \emph{Proceedings of the
  IEEE/CVF Conference on Computer Vision and Pattern Recognition}, 2020, pp.
  12\,465--12\,474.

\bibitem{leon2021maas}
J.~Le{\'o}n-Alc{\'a}zar, F.~C. Heilbron, A.~Thabet, and B.~Ghanem, ``Maas:
  Multi-modal assignation for active speaker detection,'' \emph{arXiv preprint
  arXiv:2101.03682}, 2021.

\bibitem{tao2021someone}
R.~Tao, Z.~Pan, R.~K. Das, X.~Qian, M.~Z. Shou, and H.~Li, ``Is someone
  speaking? exploring long-term temporal features for audio-visual active
  speaker detection,'' in \emph{Proceedings of the 29th ACM International
  Conference on Multimedia}, 2021, p. 3927–3935.

\bibitem{min2022learning}
K.~Min, S.~Roy, S.~Tripathi, T.~Guha, and S.~Majumdar, ``Learning long-term
  spatial-temporal graphs for active speaker detection,'' \emph{arXiv preprint
  arXiv:2207.07783}, 2022.

\bibitem{senet}
\BIBentryALTinterwordspacing
J.~Hu, L.~Shen, and G.~Sun, ``Squeeze-and-excitation networks,'' \emph{2018
  IEEE/CVF Conference on Computer Vision and Pattern Recognition}, Jun 2018.
  [Online]. Available: \url{http://dx.doi.org/10.1109/CVPR.2018.00745}
\BIBentrySTDinterwordspacing

\bibitem{guo2016ms}
Y.~Guo, L.~Zhang, Y.~Hu, X.~He, and J.~Gao, ``Ms-celeb-1m: A dataset and
  benchmark for large-scale face recognition,'' in \emph{European conference on
  computer vision}.\hskip 1em plus 0.5em minus 0.4em\relax Springer, 2016, pp.
  87--102.

\bibitem{indefense}
\BIBentryALTinterwordspacing
J.~S. Chung, J.~Huh, S.~Mun, M.~Lee, H.-S. Heo, S.~Choe, C.~Ham, S.~Jung, B.-J.
  Lee, and I.~Han, ``In defence of metric learning for speaker recognition,''
  \emph{Interspeech 2020}, Oct 2020. [Online]. Available:
  \url{http://dx.doi.org/10.21437/interspeech.2020-1064}
\BIBentrySTDinterwordspacing

\bibitem{voxceleb2}
\BIBentryALTinterwordspacing
J.~S. Chung, A.~Nagrani, and A.~Zisserman, ``Voxceleb2: Deep speaker
  recognition,'' \emph{Interspeech 2018}, Sep 2018. [Online]. Available:
  \url{http://dx.doi.org/10.21437/Interspeech.2018-1929}
\BIBentrySTDinterwordspacing

\bibitem{hoover2018using}
K.~Hoover, S.~Chaudhuri, C.~Pantofaru, I.~Sturdy, and M.~Slaney, ``Using
  audio-visual information to understand speaker activity: Tracking active
  speakers on and off screen,'' in \emph{2018 IEEE International Conference on
  Acoustics, Speech and Signal Processing (ICASSP)}.\hskip 1em plus 0.5em minus
  0.4em\relax IEEE, 2018, pp. 6558--6562.

\bibitem{sharma2022unsupervised}
R.~Sharma and S.~Narayanan, ``Unsupervised active speaker detection in media
  content using cross-modal information,'' \emph{arXiv preprint
  arXiv:2209.11896}, 2022.

\bibitem{resnet}
\BIBentryALTinterwordspacing
K.~He, X.~Zhang, S.~Ren, and J.~Sun, ``Deep residual learning for image
  recognition,'' 2015. [Online]. Available:
  \url{https://arxiv.org/abs/1512.03385}
\BIBentrySTDinterwordspacing

\bibitem{pyannote}
H.~{Bredin}, R.~{Yin}, J.~M. {Coria}, G.~{Gelly}, P.~{Korshunov},
  M.~{Lavechin}, D.~{Fustes}, H.~{Titeux}, W.~{Bouaziz}, and M.-P. {Gill},
  ``{pyannote.audio: neural building blocks for speaker diarization},'' in
  \emph{ICASSP 2020, IEEE International Conference on Acoustics, Speech, and
  Signal Processing}, 2020.

\bibitem{pardo2021moviecuts}
A.~Pardo, F.~C. Heilbron, J.~L. Alc{\'a}zar, A.~Thabet, and B.~Ghanem,
  ``Moviecuts: A new dataset and benchmark for cut type recognition,''
  \emph{arXiv preprint arXiv:2109.05569}, 2021.

\bibitem{deng2020retinaface}
J.~Deng, J.~Guo, E.~Ververas, I.~Kotsia, and S.~Zafeiriou, ``Retinaface:
  Single-shot multi-level face localisation in the wild,'' in \emph{Proceedings
  of the IEEE/CVF conference on computer vision and pattern recognition}, 2020,
  pp. 5203--5212.

\bibitem{Bewley2016_sort}
A.~Bewley, Z.~Ge, L.~Ott, F.~Ramos, and B.~Upcroft, ``Simple online and
  realtime tracking,'' in \emph{2016 IEEE International Conference on Image
  Processing (ICIP)}, 2016, pp. 3464--3468.

\bibitem{talknet}
\BIBentryALTinterwordspacing
R.~Tao, Z.~Pan, R.~K. Das, X.~Qian, M.~Z. Shou, and H.~Li, ``Is someone
  speaking?'' \emph{Proceedings of the 29th ACM International Conference on
  Multimedia}, Oct 2021. [Online]. Available:
  \url{http://dx.doi.org/10.1145/3474085.3475587}
\BIBentrySTDinterwordspacing

\bibitem{shahid_columbia}
M.~Shahid, C.~Beyan, and V.~Murino, \emph{Comparisons of Visual Activity
  Primitives for Voice Activity Detection}, 09 2019, pp. 48--59.

\bibitem{svvad}
------, ``S-vvad: Visual voice activity detection by motion segmentation,'' in
  \emph{2021 IEEE Winter Conference on Applications of Computer Vision (WACV)},
  2021, pp. 2331--2340.

\bibitem{realvad}
C.~Beyan, M.~Shahid, and V.~Murino, ``Realvad: A real-world dataset and a
  method for voice activity detection by body motion analysis,'' \emph{IEEE
  Transactions on Multimedia}, vol.~23, pp. 2071--2085, 2021.

\end{thebibliography}
